Magnetic slip casting: a review of current achievements and issues

Hortense Le Ferrand*[1,2]

[1]School of Mechanical and Aerospace Engineering, Nanyang Technological University, 50 Nanyang avenue, Singapore 639798

[2]School of Materials Science and Engineering, Nanyang Technological University, 50 Nanyang avenue, Singapore 639798

*Corresponding email: hortense@ntu.edu.sg

*Ceramic materials are ubiquitous in technologies operating under high mechanical, thermal or chemical constrains. Research in ceramic processing aims at creating ceramics with properties that are still challenging to obtain, such as toughness, transparency, conductivity, among others. Magnetic slip casting is a ceramic process where an external magnetic field is used to align the ceramic grains along preferential crystallographic directions, thereby creating controlled texture. Over the past 20 years of research on magnetic slip casting, ceramics of multiple chemistry were found to exhibit enhanced properties as a result from the texturation. This paper reviews the progress in the field of magnetic slip casting, details the processing parameters, the textures obtained for a diverse range of ceramic materials. The achieved mechanical and functional properties of the magnetically textured parts are presented. This overview of the magnetic slip casting process allows to identify critical directions for future advancement in advanced technical ceramics.*

Keywords: magnetic, texture, slip casting, ceramics, enhanced properties

**1| INTRODUCTION**

Ceramic materials are ubiquitous: dental crowns, implants, tiles, sanitary wares, thermocouples, spark plugs... Advances in ceramic formulation and their manufacturing processes have propelled traditional ceramic materials from the household to high tech industrial applications. In the past 40 years, researchers have put efforts in optimizing those processes and adapting them to a wide range of ceramic materials and applications. Those processes are tape casting [1], slip casting [2], freeze casting [3], gel casting [4] and recently, 3 dimensional (3D) printing **(Figure 1A)** [5]. One of the major challenges that this research tackles is the reduction of flaws that control the minimum stress to fracture following Griffith's criterion [6]. In addition, the optimization of chemical formulations and of their processes could generate ceramics with unique structural and functional properties, such as shape-memory and ductility [7,8], transparency [9], high capacity [10], biocompatibility [11], etc. Such properties make ceramic materials applicable in a large range of applications, from sensors [12] to actuators [13], batteries [14], implants [15], protective shields [16], reflectors [17], heat exchangers [18], etc.

Despite the outstanding results achieved so far, challenges remain, such as scale-up, obtention of near net-shape and properties like toughness, combinations of mechanical and optical functions, or thermal and electrical conductivities high enough for miniaturized electronics. Scaling-up of near net-shape manufacturing methods of ceramic parts have been investigated but tackle porous foams without those properties [20]. Near net-shape fabrication using 3D printing is currently actively researched. Here, this review tackles one of the most commonly used method to fabricate large scale ceramic parts, that is slip casting. It is a method by which a colloidal suspension is poured into a porous mold, dried, and consolidated by sintering [2]. Since it is usually not desirable for an industry to drastically change its production system, and ideal manner to improve an established process is to augment it with the addition of an equipment or processing step. For example, pressing and centrifuging during the casting have been used [21,22]. Over the past ca. 20 years, slip casting in external magnetic fields has been explored **(Figure 1B)**. The process of magnetic slip casting and the resulting ceramics are discussed in this review.

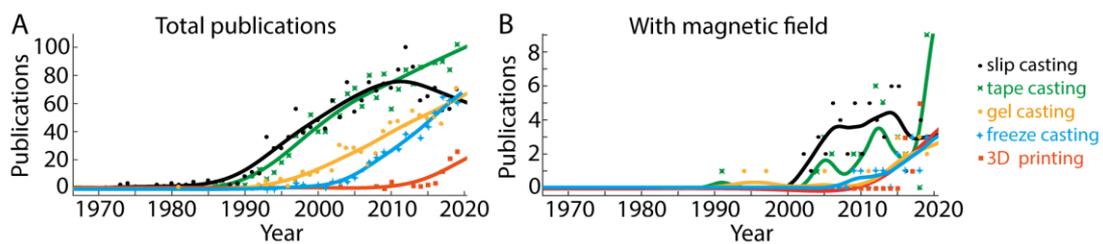

**Figure 1:** Number of publications as a function of the years on slip casting, tape casting, freeze casting and 3D printing of ceramics, without **(A)** and with external magnetic field **(B)**. Data points obtained through Isi Web of Science.

Magnetic fields are naturally occurring on the Earth, can be engineered through permanent ferrimagnets and rare-earth magnets, or can be created by electrical currents as in electromagnets (**Figure 2**). Contrary to electric fields, magnetic fields are remote, and can range over ca. 6 orders of magnitude depending on the source.

Magnetic fields as low as the Earth's magnetic field are found to induce orientation and alignment in natural ferromagnetic minerals and are used by animals for sensing and navigation [23–25]. Magnetic fields of higher strength provide the opportunity to orient ferromagnetic as well as dielectric matter in specific directions. This can be used to create anisotropy and to enhance properties along the direction of alignment. External magnetic fields have been applied during metallurgic processes during cold rolling, annealing, or crystallization, and resulted in oriented textures [26]. In ceramic processing, magnetic fields are also an interesting means to control the texture.

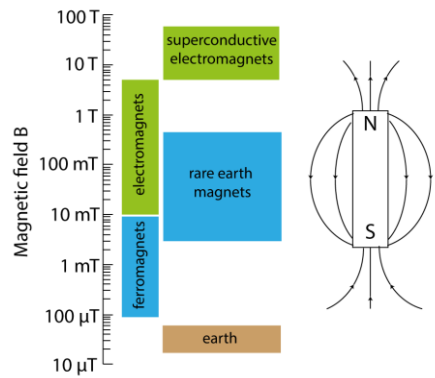

**Figure 2:** Magnetic field strength and lines.

Although a few reviews have tackled magnetic processing [27–29], there is no published review article that describes, discusses and deciphers the potentials and limits of magnetic slip casting, specifically. Although 62 research publications on magnetic slip casting applied to ca. 30 ceramic compositions can be found **(Table 1)**, this review is timely as a new trend is emerging. Indeed, ultralow magnetic fields of a few mT only are now being investigated instead of the high strength ones of 10-12 T. Using ultra low magnetic fields to create orientation was first exploited in soft composites [30,31] and is now investigated for pure ceramics [32,33]. In this paper, the general principles of slip casting and following sintering are first described. Then, the requirements for the magnetic orientation of ceramic powders are reviewed as well as the magnetic set-ups used. Following are detailed the characterization of magnetically oriented textures and possible causes for defects and misalignments. Finally, properties and applications are reviewed and their limits and potential highlighted. By providing this overview of magnetic slip casting, researchers from different areas such as materials science, electromagnetics, optics, as well as engineers and industrials, could work together to further enhance the properties and applications of ceramics using magnetically controlled texturation.

**Table 1:** Number of publications to date on different ceramic processes, without and with magnetic field, as referenced in Isi Web of Science.

| | Total number of publications to date | |
|---|---|---|
| Name of the process | Without B | With B |
| Slip casting | 1719 | 62 |
| Tape casting | 1692 | 38 |
| Gel casting | 821 | 16 |
| Freeze casting | 584 | 14 |
| 3D Printing | 81 | 15 |

## 2| SLIP CASTING OF CERAMICS
### 2.1| Principle

In the following, the principle of slip casting and its mechanisms are described. Slip casting is a colloidal ceramic process by which a concentrated stable suspension, the slip, is casted onto a porous mold. The slip should be a homogeneous and well dispersed suspension. Indeed, the formation of agglomerates or sediments would lead to anisotropic shrinkage leading to cracks, warping and deformation at the drying and sintering stages. The porous mold is traditionally made of gypsum, or plaster of Paris. However, other porous molds can be used, such as polymeric or alumina molds with activated carbon [34]. Adding 250 nm diameter activated carbon to the alumina mold increases its permeability in a controlled fashion, making those molds sometimes preferable to gypsum. Also, gypsum leaves traces of calcium and sulfur at the surface of the casted body. Alumina molds reduce the surface contamination, which can be critical for example, for transparent ceramics [35].

During the casting, the capillary pressure from the porous mold induces a flow that slowly removes the solvent from the suspension, depositing the ceramic particles from the suspension at the surface of the mold. The kinetics of the solvent removal is typically described by Darcy's law where the flux $J$ of the filtrated fluid is [35]:

$$J = \frac{dV}{A\,dt} = \frac{K \cdot \Delta P}{\eta \cdot H},$$  (**equation 1**)

where $K$ is the permeability of the porous deposit, $\eta$ the viscosity of the solvent, $\Delta P$ the pressure difference across the porous deposit, $t$ the casting time, $H$ the thickness of the porous deposit and $V$ and $A$ its volume and area, respectively. Deriving equation 1 gives the growth of the deposited layer as a function of time:

$$H = \sqrt{\frac{K \cdot \Delta P}{\eta} t}.$$  (**equation 2**)

After total suction of the solvent, the deposited particles assembly forms a wet green body that is subsequently dried. This green body is then sintered to yield a ceramic. The sintering step may be preceded by a pressing step to further increase the packing of the ceramic particles. To allow the handling of the dry green body and reduce cracking during drying, a polymeric additive, the binder, is usually added to the suspension. The binder adds plasticity to the wet green body and gives strength to the dry green body. Considering the case of spherical particles, the mechanical strength $\sigma$ of the green body depends on the binder concentration [36]:

$$\sigma = \frac{9\phi}{8\sqrt{1-\sigma}}\sqrt{v_B}\sigma_B,\qquad\text{(equation 3)}$$

where $\phi$ is the packing fraction of the ceramic particles, $v_B$ the volume fraction of binder relative to the ceramic and $\sigma_B$ the strength of the binder. $\sigma_B$ represents either the polymer's cohesive strength or the adhesion forces at the polymer-ceramic interface. Before sintering, the binder should be burnt out without the creation of cracks or other deformation. Thanks to the sacrificial nature of the binder, it can be used to create dense ceramics [36], or to create well defined internal pores or channels [37].

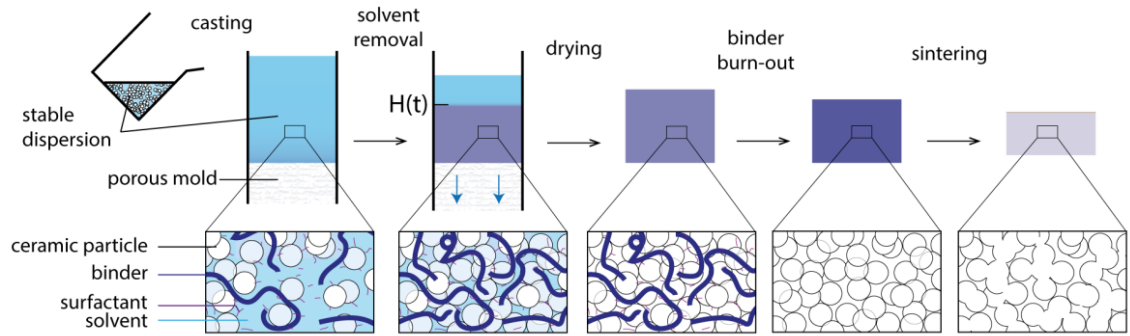

**Figure 3:** Schematics of the slip casting process.

## 2.2| Sintering of slip casted parts

Sintering needs a specific description of its own since the mechanisms involved are independent of the casting. During the sintering, the densification of the green bodies into ceramic parts occurs. Necks develop between the ceramic particles following a diffusion process. Conventional sintering uses thermal activation of the mass transport at temperatures ½ or ¾ of the melting temperature. To reduce the time, temperature, and energy consumption associated with sintering, a variety of methods have been developed [38]. First, pressure assisted sintering such as unidirectional pressing and hot isostatic pressing have been used. Pressure acts on the sintering by increasing the chemical potential of the ceramic particles and promoting atomic diffusion. Applied pressure can also deform the particles *via* viscous or plastic flows and creep, as well as increase the particles packing and arrangement while destroying agglomerates [39]. Pressure can be applied in concurrence with electric stimulation. This is referred as spark plasma sintering (SPS), pulsed electric current sintering (PECS), field assisted sintering technique (FAST) or current activated pressure assisted densification (CAPAD) [40]. Finally, sintering at temperatures lower than 200 °C can be realized by combination of hydrothermal processing and isostatic pressing, but the mechanisms are still not completely understood [41].

However, application of pressure may limit the shaping of the ceramics, that is one key advantage of the slip casting process. Indeed, in slip casting, the shape of the part is controlled by the shape of the mold, enabling the creation of curved, elongated or complex shapes. To remedy to this issue, pressure-less sintering is applied in presence of additives. These additives increase the rate of densification and decrease the required temperature without compromising the final density. For example, liquid phase sintering densifies ceramic with a glassy phase at the grain interfaces [42]. As the sample is heated at the melting temperature of the glass, capillary forces pull the ceramic grains together while the diffusion is increased in the viscous layer between the grains. Another method uses impurities to promote the migration of the grain boundaries and the epitaxial growth of grains [43]. This abnormal grain growth is realized by distributing larger particles in a fine-grained matrix. When the large particles are anisotropic with well-defined crystallographic orientations, the process is referred to as templated grain growth [44]. Using microwaves for heating has also been found to enhance diffusion by heating the ceramic body homogeneously through electromagnetic energy absorption [45].

The determination of the optimum sintering parameters depends on many factors, such as powder size and composition, green body density, sintering additives, and are usually determined using dilatometry. Computational modeling and simulations may help optimize the process [46]. Artificial intelligence and machine learning also open interesting means to optimize the sintering to achieve desired properties [47].

Slip casting is thus a process in two steps. To create magnetically oriented textures using slip casting, the external magnetic field is applied during the casting, when the slurry is dispersed. The magnetic orientation process is described in the following section.

## 3| MAGNETIC ORIENTATION
### 3.1| Magnetic anisotropy

Magnetic slip casting consists in the application of an external magnetic field during the casting process to orient preferential crystalline axis of the ceramic grains. Most ceramic phases, other than ferrites and spinel, are diamagnetic. Applying an external magnetic field to a diamagnetic material induces an internal magnetic field in the other direction, causing a repulsive force. Generally, to create such induced internal magnetic field, the strength of the applied field needs to be very high because the magnetic susceptibility of diamagnetic materials is weak. Transforming the diamagnetic ceramic powder using magnetic elements is a way to increase its magnetic susceptibility and lower the external magnetic field strength required. The orientation of the ceramic powder results from the crystalline or shape magnetic anisotropy of the starting ceramic powder (**Figure 4**):

Crystalline anisotropy is an intrinsic property of some crystal structures where the atomic arrangement and the deformation of the atomic electron clouds lead to magnetic anisotropy. It is observed that even diamagnetic crystals such as most ceramic crystals display magnetic orientation along a preferred axis, when the magnetic field strength is high enough [48]. This orientation arises from anisotropy in the magnetic susceptibility due to electron orbits that cover different areas along different directions in the crystal [49]. The anisotropy in magnetic susceptibility $\Delta\chi$ between the directions parallel ($\chi_\parallel$) and perpendicular ($\chi_\perp$) to the c-axis has been measured using field-induced harmonic oscillation, where the moment of inertia $I$ of a crystalline rod is recorded as the magnetic field strength $B$ increases [50]. In the high field region (10 T), harmonic oscillations are observed and their period $\tau$ measured, giving:

$$\Delta\chi = \frac{4\pi^2 I}{(\tau B)^2 N},$$ (**equation 4**)

with $N$ the mole number of the crystal rod and $l$ its length. For example, for $Al_2O_3$, $\Delta\chi \sim 7 \times 10^{-10}\ emu/g$ [51]. As a comparison, for antiferromagnetic magnetite $Fe_2O_3$, $\Delta\chi \sim 3.8\ emu/g$ [52]. The free energy of magnetization U is then varying depending on the direction:

$$U = -\frac{\chi}{2\mu_0(1+M\chi)^2},$$ (**equation 5**)

where $\mu_0$ is the vacuum permeability and $M$ the demagnetization factor, and $\chi$ either $\chi_\parallel$ or $\chi_\perp$. Generally, in the case where $\chi_\parallel > \chi_\perp$, $U(\chi_\parallel) < U(\chi_\perp)$, leading to the preferred orientation along the parallel direction to minimize the energy. This preferential orientation generates a torque $T$:

$$T = \frac{1}{2\mu_0} V \Delta\chi B^2 \sin 2\theta,$$ (**equation 6**)

with $V$ the volume and $\theta$ the angle between the axis of highest susceptibility and the imposed magnetic field [53].

Shape anisotropy, on the contrary, is the magnetic orientation of crystalline particles that display an anisotropic geometry such as rod or disc, but an isotropic susceptibility. It is the shape of the particle that determines the anisotropy in susceptibility and the preferential orientation. Typically, for a spheroid of bulk magnetic susceptibility $\chi$, the intensity of magnetization $J$ under an external field $B$ is:

$$J = \chi_a B = \chi(B - M \cdot J),\qquad\text{(equation 7)}$$

leading to:

$$\chi_a = \frac{\chi}{1+M\chi},\qquad\text{(equation 8)}$$

with $\chi_a$ the apparent magnetic susceptibility [52]. $\chi_a$ is a second order symmetric tensor. In the case of an ellipsoid, $\chi_{aa} = \chi_1 \neq \chi_{ab} = \chi_{ac} = \chi_2$. The ratio of magnetic susceptibilities is then:

$$\frac{\chi_1}{\chi_2} = \frac{1+M_2\chi}{1+M_1\chi},\qquad\text{(equation 9)}$$

with $M_1$ and $M_2$ the diamagnetizing factors along the two main directions. As a consequence, shape anisotropy is large only if the isotropic magnetic susceptibility $\chi$ is high enough. In addition, it was measured that although the anisotropy increases with the aspect ratio of the crystal, it plateaus above an aspect ratio of ~100 [52].

To use shape magnetic anisotropy for aligning diamagnetic ceramics, there is a need to increase the bulk susceptibility $\chi$ of the anisotropic crystals. This can be achieved by coating the powder with ferri-, ferro- or para-magnetic elements, attaching superparamagnetic nanoparticles at their surface, coating magnetic particles with ceramic, or doping the ceramic crystals with magnetic elements. The sol-gel coating of microplatelets of alumina with iron could increase the bulk susceptibility to $4.95\ emu/g$, close to that of magnetite, whereas decoration with superparamagnetic iron oxide could increase the susceptibility to $9.89\ emu/g$, three times higher than magnetite [54]. Preparation of spherical nanoparticles by sequential synthesis yielded iron-core alumina spheres with layers of $\gamma$-$Fe_2O_3$/$SiO_2$/$\gamma$-$Al_2O_3$ with magnetic susceptibility of $3.77 \times 10^{-5}\ emu/g$, 5 orders of magnitude higher than alumina [55]. Finally, computational modeling show that doping of alumina with Fe, Ni, Co, Mn, etc. creates spin delocalization that could generate a magnetic response at low field strength [56].

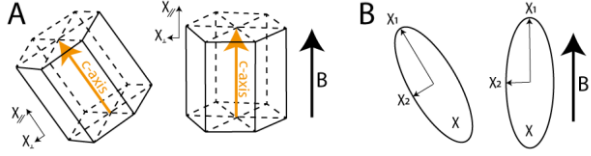

**Figure 4**: Schematic representation of magnetic alignment due to crystalline anisotropy **(A)** and shape anisotropy **(B).**

**3.2| Particle size effect**

In addition to the anisotropic magnetic susceptibility, the size of the individual particles to align with the magnetic field is critical to determine the processing parameters, such as field strength and alignment time. Indeed, there are two major opposing forces that need to be considered, namely the Brownian motion and the sedimentation under gravity. For spherical particles of radius $r$ and density $\rho_p$ suspended in a liquid of density $\rho_l$, three radii have been derived [57]:

$$r_{max} = \sqrt{\frac{3 \cdot \Delta\chi \cdot B^2 \cdot L}{4 \cdot \mu_0 \cdot g \cdot (\rho_p - \rho_l)}}, \quad \text{(equation 10)}$$

$$r_{crit} = \sqrt[7]{\frac{243 \cdot T_{mp} \cdot \eta^2 \cdot k_B}{8 \cdot \pi \cdot e \cdot g^2 \cdot \rho_p \cdot (\rho_p - \rho_l)^2}}, \quad \text{(equation 11)}$$

$$r_{min} = \sqrt[3]{\frac{3 \cdot T_{mp} \cdot \mu_0 \cdot k_B}{2 \cdot \pi \cdot \Delta\chi \cdot B^2}}, \quad \text{(equation 12)}$$

where $L$ is the thickness of the cast, $T_{mp}$ the temperature, $\eta$ the viscosity of the liquid, $g$ the acceleration of gravity and $k_B$ Boltzmann constant. These radii describe that for $r_{min} < r < r_{crit}$, the particle is small enough to neglect the effect of gravity, whereas the Brownian motion has to be overcome; for $r_{crit} < r < r_{max}$, the particle is large enough to neglect Brownian motion but the gravitational effects have to be overcome. It thus exists a critical radius at which both Brownian motion and gravity are negligible as compared to the magnetic effect, *i.e.* for $r \approx r_{crit}$. At this critical dimension, the particles exhibit a so-called ultra-high magnetic response (UHMR). Typically, for $Al_2O_3$ particles, $r_{crit} \approx 5 - 50 \, \mu m$ [31].

Finally, in the case $r > r_{max}$, it was derived that for ellipsoid particles of aspect ratio $l/2r$, the particles can align if the adimensional coefficient $C = \frac{\Delta\chi \cdot B^2}{\mu_0 \cdot g \cdot \rho_p \cdot 2r}$ is large enough. In addition, if $\Delta\chi > 0$, a small aspect ratio is preferred for magnetic alignment under a field parallel to the gravity direction, whereas if $\Delta\chi < 0$, a large aspect ratio will be favorable for alignment under a field perpendicular to the gravity direction [57].

### 3.3| Magnetic set-ups

The magnetic set-ups used in magnetic slip casting can be classified in two groups: those using electromagnets and those using permanent magnets. In both cases, the magnetic field or the sample can be rotated. Rotation is often favorable for magnetic alignment, along a single direction but for biaxial orientation. Indeed, during rotation, the particles undergo a viscous drag. There is an optimum rotation frequency at which ellipsoid particles align along the rotating magnetic field, at equilibrium [58]. Ellipsoids of volume $V$ suspended in a liquid of magnetic susceptibility $\mu_l$ and submitted to a precessing magnetic field $\boldsymbol{B}$ undergo a viscous torque $\boldsymbol{T_v} = -\varsigma\boldsymbol{\omega}$ with $\varsigma$ the rotational friction coefficient, and a magnetic torque $\boldsymbol{T_m} = \mu_l \cdot V \cdot \Delta\chi \cdot (\boldsymbol{n} \times \boldsymbol{B})(\boldsymbol{n} \cdot \boldsymbol{B})$ with $\boldsymbol{n}$ the preferred orientation axis of the ellipsoid. Solving the equilibrium equation between the two torques yields a critical frequency above which the particle aligns along $\boldsymbol{B}$ [58]:

$$\omega_c = \mu_0 \frac{V \cdot \Delta\chi \cdot B^2}{2\varsigma} \quad \textbf{(equation 13)}.$$

The same description is valid for crystals with crystalline magnetic anisotropy.

**Table 2** summarizes the types of set-ups used in magnetic slip casting of various ceramics and their related advantages and limitations. Electromagnets are coiled wires where a voltage is applied to generate a high current that will create the magnetic field following Ampere's law. The magnetic field lines inside the cavity of the coil are parallel to the edges of the electromagnet. When a supraconducting wire is used to conduct the current, the magnetic field strength can be increased up to 45 T [59] (**Figure 2**). With such a high magnetic field, electromagnets have been used to align diamagnetic particles of feeble magnetic susceptibility. A rotating platform can be added to support the sample and create rotating magnetic field by rotating the sample. Permanent magnets, on the contrary, generate lower magnetic field, up to 500 mT for standard large NdFeB magnets (**Figure 2**). Magnetic cubes larger than 20 cm width would create larger magnetic fields but the attracting force would pause safety issues. Despite a low and less homogeneous magnetic field, permanent magnets have been effectively used to orient crystals and particles that have a high magnetic susceptibility. In addition, their small size makes them easy to handle. Small magnets have been used in static and rotating systems to orient particles in ceramics in any directions in space by sticking magnets to multiple motors [60,61].

**Table 2:** Magnetic set ups

| Type of set-up | Magnetic field range | Advantages | Limitations | | Publications |
|---|---|---|---|---|---|
| Electromagnet | 3-12 T | Homogeneous field at the center of the electromagnet, high strength magnetic field, can be turned on and off | Bulky equipment, high electrical current inputs, sample dimensions limited by the core of the magnet, room temperature only. | Static horizontal | BaTiO$_3$ [62], ZrB$_2$ [63], HA [64,65], LiCoO$_2$ [66,67], Bi$_2$Te$_3$ [68] |
| | | | | Static vertical | Bi$_4$Ti$_3$O$_{12}$ [57], Si$_3$N$_4$ [69–71], Y$_2$O$_3$-AlN [72], TiB$_2$ [73,74], MgB$_2$ [75], CeF$_3$ [76], TiBaSrCa$_2$Cu$_3$O$_9$ [77], LiNbTiO [78], 3Y-TZP [79], TiO$_2$ [80,81], SiC [82,83], Al$_2$O$_3$ [84–89], Nb$_4$AlC$_3$ [90], Pb(Zr,Ti)O$_3$ [91], ZnO [92], BaTiO$_3$ [93], BNIT [94], $\beta$-TCP [95], Ti$_3$SiC$_2$ [96], AlN [97], CaSi$_3$N$_4$ [98] |
| | | | | Rotating horizontal | B$_4$C [99], zeolite [100], Si$_3$N$_4$ [101,102], ErBa$_2$Cu$_4$O$_8$ [103], HA [104,105], $\beta$-TCP [95], (Ca,Sr)Bi$_4$Ti$_4$O$_{15}$ [106] |
| | | | | Rotating vertical | Bi$_{0.5}$Sb$_{1.5}$Te$_3$ [107], HA [108], Ti$_3$SiC$_2$ [96,109,110] |
| Rare-earth magnet | 0.01-0.5 T | Cheap, small equipment, various shapes and assemblies possible, any field orientation possible. | High inhomogeneities, sample dimensions limited by magnet size and strength, can be used up to the Curie temperature (ca 60 °C) | Rotating in any directions | Al$_2$O$_3$ [61,111–113] |

## 4| MAGNETICALLY ORIENTED TEXTURES
### 4.1| Characterization methods of the texture

Magnetic fields orient particles during casting. After sintering, the ceramic exhibits a texture, hence a narrow distribution of crystalline orientations along one direction. The texture can be characterized using several methods that make use of the crystalline anisotropy in the ceramic, or the shape anisotropy of its grains. Crystalline anisotropy is typically recorded using the diffraction of an incoming wave onto the crystallographic planes following Bragg's law, using X-rays or neutron rays. Shape anisotropy can be measured using X-ray scattering under small angles (SAXS), through direct imaging of the grain shape by electron microscopy, or by measuring the light reflection at the surface of the ceramic. Some characteristics of those methods, the orientation parameters and relevant references relative to their use for characterizing samples prepared by magnetic slip casting are summarized in **Table 3**.

**Table 3**: Methods used for characterizing the texture of ceramics prepared by magnetic slip casing.

| Method | Type of anisotropy | Sample preparation | Probed area/volume; resolution | Orientation parameter | Ref. |
|---|---|---|---|---|---|
| Selected Area Electron Diffraction (SAED) in Transmission Electron Microscope (TEM) | Crystal | Ultrathin sample lamellae | ~10-100 µm², atomic resolution | Tilting angle from the main axis: $\varphi = \cos^{-1}(\cos x \cdot \cos y)$ Direction of tilting in the plane: $\theta = \tan^{-1}(\frac{\tan y}{\sin x})$ | [78] |
| Small Angle X Ray Scattering (SAXS) | Crystal & shape | Polished sample, thin lamellae | ~1-3 mm³ | Orientation parameter $S$ $S = \frac{\int_{\theta=0}^{2\pi} I(\theta)M(\theta)d\theta}{\int_{\theta=0}^{2\pi} I(\theta)d\theta}$ | [113] |
| Electron Backscattered Diffraction (EBSD) in Scanning Electron Microscope (SEM) | Crystal | Atomically polished surface | ~10-1000 µm², 100 nm resolution | Orientation parameter $r$ $f_{MD}(r,\theta) = (r^2 \cos^2\theta + \frac{\sin^2\theta}{r})^{-3/2}$ | [66,67,80,83,85,97] |
| X-Ray Diffraction (XRD) $\theta - 2\theta$ scan | Crystal | Polished surface | ~1-3 mm³ | Lotgering factor $f = \frac{p-p_0}{1-p_0}$, $p = \frac{\sum I(00l)}{\sum I(hkl)}$ and $p_0 = \frac{\sum I_0(00l)}{\sum I_0(hkl)}$ | [73,74,79,99,110] |
| X-Ray Diffraction (XRD) Rocking curves | Crystal | Polished surface | ~1-3 mm³ | Full Width at Half Max (FWHM) | [113] |
| Scanning Electron Microscopy (SEM) | Shape | Fractured surfaces or polished and etched surfaces | ~1-100 µm², 100 nm resolution | Orientation parameter $S$ $S = \frac{1}{2}(\cos^2\theta - 1)$ | [113] |
| Neutron diffraction $\theta - 2\theta$ scan | Crystal | Bulk parallelepiped shape | ~3.2 cm³ | Orientation parameter $r$ $\frac{I_{obs}(hlk)}{I_{cal}(hlk)} = s \cdot (r^2 \cos^2\theta + \frac{\sin^2\theta}{r})^{-3/2}$ | [85] |
| Directional Reflectance Microscopy (DRM) | Shape | Polished surfaces | ~1-100 cm², 5 µm resolution | Reflection angle distribution | [60] |

Texture on the surface of ceramics can be determined using X ray emissions in electron microscopes such as electron backscattering diffraction (EBSD) in scanning electron microscopes (SEM) [66,67,80,83,85,97] or selected area electron diffraction (SAED) in transmission electron microscopes (TEM) [78]. Although those two methods provide a nanometric resolution and the characterization of individual grains, the probed area is limited to a few µm. The output data can be presented under the form of diffractions peaks for SAED and pole figures for EBSD, from which orientation parameters can be derived (**Table 3**). To allow for large area scanning, X ray diffraction (XRD) in diffractometers can probe mm dimensions and with a penetration depth of a few hundreds of micrometers. XRD can be carried out using the $\theta - 2\theta$ scan method as in powder diffraction [73,74,79,99,110] or using the rocking curve method where only the diffraction around one preferential axis is recorded by tilting the sample [113]. One advantage of using rocking curves over $\theta - 2\theta$ scans is the direct determination of the misalignment degree by measuring the full width at half maximum (FWHM) of the scattered intensity at the preferential orientation. Alignment of anisotropic crystals can also be visualized at the microscale using SEM imaging and image analysis [113]. However, surface methods provide information on the orientation of particles only in one plane, requiring multiple views for determining the 3D orientation throughout bulk samples.

Neutron diffraction and small angle X ray scattering (SAXS) under synchrotron radiation have been used to gather information in bulk specimen of a few millimeter thickness and width [85,113]. Whereas SAXS provides information on both crystalline and shape anisotropy, neutron scattering determines the texture from a crystalline point of view only. Thanks to a higher penetration depth, neutron scattering allows the characterization of texture throughout samples up to a few centimeters in thickness. Combining neutron diffraction with synchrotron X-ray microcomputed tomography provides the opportunity to measure the texture and the grain dimensions in 3D in bulk specimens [114].

Finally, to probe macroscopic samples at large scales and short times, light scattering can be used. Indeed, light will exhibit different scattering intensity depending on the surface topography of a sample. This surface topography can be revealed by fine polishing of surfaces combined with etching methods and is referred to as directional reflectance microscopy (DRM) [115]. In textured ceramics with shape anisotropy, it is hypothesized that the change in light reflection with grain orientation is due to the number of interfaces between the grains as well as the dimension of the grain [60,113]. In textured ceramics with crystalline anisotropy, it is the etching protocol that would induce a light reflection correlated to the crystal structure, as in metal specimen [115]. However, the method has not been applied yet to etched ceramics.

**4.2| Textures obtained**

The textures of the ceramics obtained by magnetic slip casting are generally oriented in one direction that is parallel or perpendicular to the horizontal. Thanks to the use of permanent magnets and lower magnetic fields, it has been recently possible to orient the textures in any direction, as well as creating ceramics with multiple local controlled textures [60,113]. In the following are compared textures that were characterized using the Lotgering factor $f$. Indeed, this is by far the factor the most used in the literature to characterize textured ceramic (**Figure 5**). The closer to 1 $f$ is, the highest is the texturation, that is, the alignment. After magnetic slip casting and sintering, the ceramics obtained exhibited a high degree of orientation, with $f$ generally ranging well above 0.8 (**Figure 5A**). Some studies reported lower values, but it was explained by the authors that agglomerates inhibited the orientation of the ceramic powder during the process, leading to only $f = 0.43$ [62]. A high Lotgering factor confirms a high alignment and texturation and is found to be independent to the average magnetic susceptibility of the composition considered. Examples of alignment direction obtained for a variety of ceramics are given in **table 4**.

**Table 4:** Main axis alignment in most common ceramics using crystalline anisotropy

| Chemistry | Target properties | Crystal system | Alignment | Ref. |
|---|---|---|---|---|
| TiB$_2$ | Thermoelectric | hexagonal | c-axis parallel to B | [73,74] |
| Al$_2$O$_3$ | Mechanics, transparency | hexagonal | c-axis parallel to B | [86–88,116] |
| Ti$_3$SiC$_2$ | Mechanics, Thermal conductivity | hexagonal | c-axis perpendicular to B | [96,109,110] |
| BaTiO$_3$ | Piezoelectric | tetragonal | c-axis parallel to B | [62,93,117] |
| SiC | Mechanics | hexagonal | c-axis parallel to B | [82,83] |
| Si$_3$N$_4$ | Thermal conductivity | hexagonal | c-axis perpendicular to B | [70,71,101,102] |
| Ca$_5$(PO$_4$)$_3$(OH) (hydroxyapatite) | Mechanics, Biocompatibility | hexagonal | c-axis perpendicular to B | [65,104,105,108] |
| Ca$_3$(PO$_4$)$_2$ (ß-TCP) | Mechanics, Biocompatibility | rhombohedral | c-axis parallel to B | [95] |
| TiO$_2$ | Mechanics | tetragonal | c-axis parallel to B | [81] |

Furthermore, it is generally observed that the alignment increases with the applied magnetic field (**Figure 5B**), although the minimum field to reach $f = 1$ varies with the chemical composition and the susceptibility of the material. This increase of alignment with B follows an exponential trend (**Figure 5C**).

After magnetic orientation during the casting, the consolidated green body is sintered into a ceramic. Although the sintering parameters, heating rate, final temperature, sintering time depend on the chemistry, the degree of alignment increases during this step (**Figure 5D,E**). The increase in the Lotgering factor during sintering, despite the fact that no magnetic field is applied, is presumably due to the expitaxial growth of the grains preferentially around those that are oriented, such as minimizing the grain boundary energies. In the case of AlN ceramics, doping amounts of Y$_2$O$_3$ were added to facilitate the grain growth process and increase the degree of texturation [97]. In the case of (Ca,Sr)Bi$_4$Ti$_4$O$_{15}$, it is thought that the large oriented particles coarsen the smaller randomly oriented particles by Ostwald ripening mechanism [106]. Si$_3$N$_4$, however, might also involve a phase transformation mechanism from $\alpha$ to $\beta$ phase [70]. Indeed, $\beta - $Si$_3$N$_4$ elongated particles were aligned in a matrix of $\alpha$- Si$_3$N$_4$ before sintering, similar to the templated grain growth mechanism.

Finally, applying a rotating magnetic field is beneficial to increase the alignment, as explained earlier, although a too high-speed rotation disturbs the alignment (**Figure 5F**). This effect was observed for vertical field and horizontal rotation, as well as horizontal field and horizontal rotation, in a zeolite [100]. Rotation of the field or of the sample is particularly critical for realizing planar biaxial orientation of disc-shaped particles that exhibit shape anisotropy and that align above the critical frequency described in **equation 13** [113]. Similar energetic considerations in the case of particles displaying crystalline magnetic anisotropy explain for the positive effect of the rotation. However, too high rotation, especially if it is the sample that is rotating, might accelerate the particles, create flow and collisions, and disturb the orientation.

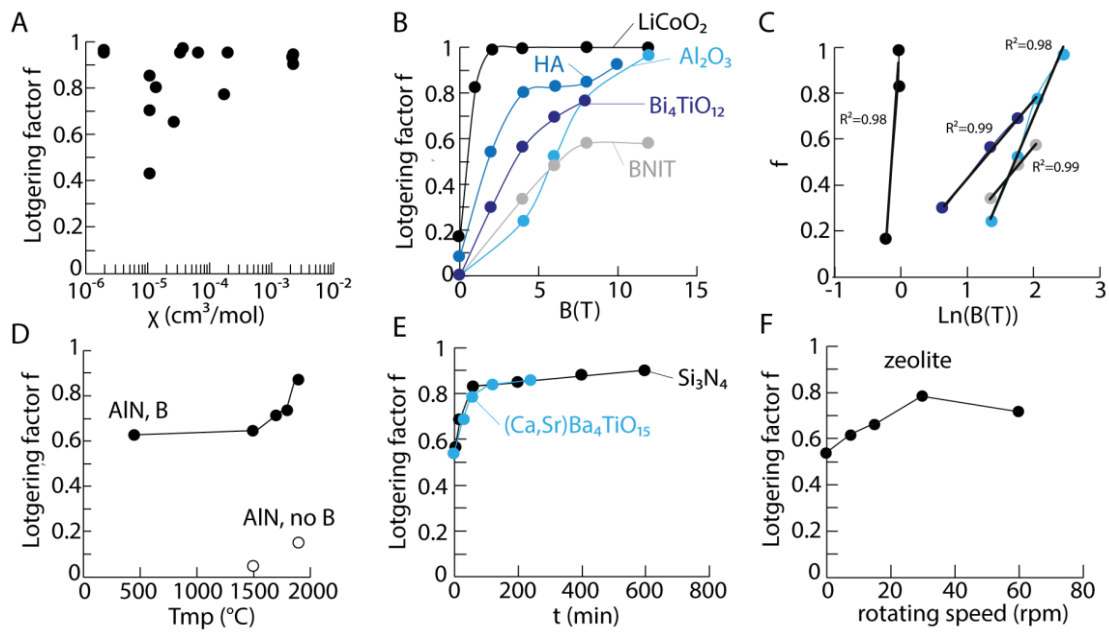

**Figure 5**: Lotgering factor $f$ as a function of **(A)** the maximum magnetic susceptibility of ceramics. Data points extracted from refs [62,63,93,99,110,117,64,66,73,74,76,79,86,92], **(B)** the applied magnetic field strength. Data points extracted from refs [66,94,118,119]. **(C)** are exponential extrapolations, **(D)** the sintering temperature. Data points extracted from ref [97], **(E)** the sintering time. Data points extracted from refs [70,106], and **(F)** the rotating speed of the sample. Data extracted from ref [100].

### 4.3| Defects and misalignment

Several defects and local misalignment may occur due to the casting process, inhomogeneities in the magnetic field and other effects from the mold design (**Figure 6**).

First, during slip casting, the solvent from the slurry is removed by capillary forces through the pores of the mold. The sucking force created generates a hydrodynamic flow at the initial stage of the process that leads to horizontal grain orientation. Indeed, in the first instants of the casting, when the powder deposit is too thin resist the flow, the hydrodynamic torque dominates the magnetic torque (**Figure 6A**) [60]. In the case of a slurry containing 25% of anisotropic alumina microplatelets, these hydrodynamic effects affect the first ~500 µm of the cast. In other compositions, such as $Bi_2Te_3$ and $Bi_4Ti_3O_{12}$, this misaligned layer reached 1.2 mm and 3.5 mm, respectively (**Figure 6B**) [57,68]. The difference in misaligned layer thickness may be due to variations in the slurry viscosity, with only 20 vol% solid loading for $Bi_4Ti_3O_{12}$, for example, or to a larger porosity of the substrate.

Secondly, the magnetic fields lines are not homogeneous in direction and intensity. Modulations in magnetic field strength can generate gradients in the slurry due to magnetophoretic forces [120–122]. It is therefore challenging for large samples to be placed in an area where the field lines are perfectly parallel. Zhu *et al.* measured the local Lotgering factor across the thickness of a sintered $Si_3N_4$ prepared under a rotating magnetic field [102]. It was observed a variation in the alignment along the thickness, with higher alignment at the center of the sample (**Figure 6C**).

Finally, the capillary forces and adhesion of the slurry to the edges of the mold can create misalignment in the diameter of the sample. Indeed, generally, a meniscus forms at the edge of the mold. In the case of a multilayer ceramic where each layer has a different orientation, this distortion is visible macroscopically thanks to different light reflections (**Figure 6D**) [61].

However, the alignment and texture achieved is far greater from non-magnetic slip casting, leading to enhanced and anisotropic structural functional properties.

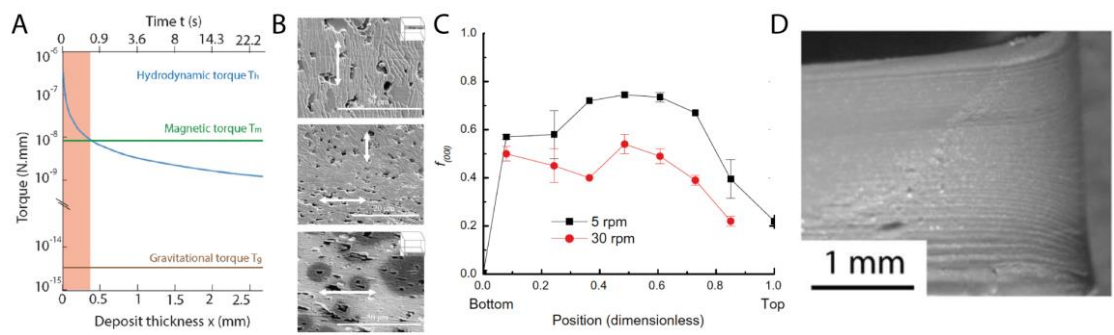

**Figure 6**: **(A)** Calculated hydrodynamic, magnetic and gravitation torques as functions of the casting time $t$ and the deposit thickness $x$ in the first instants of slip casting. Reproduced from ref [60] with permission from the Royal Society of Chemistry. **(B)** Micrographs of a cross section of a sintered sample of $Bi_4Ti_3O_{12}$ showing a horizontal

layer of 3.5 mm thickness (bottom), a middle layer with both vertical and horizontal orientation (middle), and a vertical layer (top). Reproduced with permission from [57]. Copyright 2006, Acta Materialia Inc. Published by Elsevier Ltd. **(C)** Lotgering factor of a sintered ceramic of $Si_3N_4$ at different positions parallel to the rotating field. Reproduced with permission from [102]. Copyright 2009, Acta Materialia Inc. Published by Elsevier Ltd. **(D)** Optical image of the thickness of a green body of $Al_2O_3$ with multilayer orientations showing meniscus effect. Reproduced with permission from [61]. Copyright 2015, Springer Nature.

## 5| PROPERTIES AND APPLICATIONS
### 5.1| Structural properties

Magnetic slip casting creates oriented structures. It is known that anisotropic structures in reinforced composites leads to anisotropic properties with enhanced strength and hardness along the direction of the reinforcement, whereas toughness is increased along the perpendicular direction. Similar properties are observed in textured ceramics (**Figure 7**).

Anisotropy in bending strength of textured $Ti_3SiC_2$-based ceramics has been recorded with an increase of ca. 25% in the direction parallel to the alignment (**Figure 7A**). However, it may be noted that the bending strength of ceramics of same composition critically depends on the fabrication process and the final grain size, as illustrated by the points $Ti_3SiC_2$ (1) and (2) in **Figure 7A**. Furthermore, ceramics prepared using magnetic field also exhibited an increase in toughness $K_{1c}$ up to 50% (**Figure 7B**), when the sample is loaded perpendicularly to the alignment. The parameter $K_{1c}$ represents the energy to furnish to initiate a crack. Texturing ceramics therefore provides a way to delay fracture in ceramics. Furthermore, anisotropy in $K_{1c}$ is also present, as expected. Only one paper reported anisotropic values, in a yttria-stabilized $ZrO_2$ ceramic [79] ($ZrO_2$ points in **Figure 7B**). Presumably, the intrinsic toughness of $ZrO_2$ enabled the measurement of toughness parallel to the grain orientation, otherwise very low in other ceramic formulations without intrinsic toughness. Hardness also showed anisotropy with a higher hardness along the orientation (**Figure 7C**). When the grains are aligned parallel to the field, they are generally longer in that direction too, decreasing the number of weak interfaces in this direction and increasing the hardness, as compared to the other directions. The anisotropy difference is amplified when using the templated grain growth process (TGG) for sintering, but the overall hardness values were found higher when the oriented green bodies were hot pressed (**Figure 7D**). This could be linked to the higher change in grain dimensions during TGG decreasing the anisotropy of the grains (aspect ratio ~5 after TGG as compared to ~35 after hot pressing) or to interlocking of grains due to the pressing force. Resulting from magnetic texture is also the higher wear resistance in the direction perpendicular to the orientation (**Figure 7E**).

Finally, one particularity of ceramics with aligned grains is the development of tortuous crack paths. In $Al_2O_3$ samples prepared by magnetic slip casting with an inter-grain glassy phase and sintered using hot pressing, a rising R-curve could be measured (**Figure 7F**) [61]. This rising R-curve, demonstrating the non-brittle fracture of the oriented ceramic, results from the crack deflection at the interface between elongated grains when the pre-crack and the loading are applied perpendicularly to the texture (**Figure 7G**) [123]. This phenomenon, however, depends on the sintering method and the formulation. For example, SiC textured ceramics prepared by conventional sintering exhibit a higher crack deflection and tortuosity when the sample is loaded parallel to the orientation (**Figure 7H**) [82]. In these samples, there was no glassy interfacial layer between the grains, that would have provided the interfacial strength necessary to deflect a crack at a ~80-90 degree angle [124].

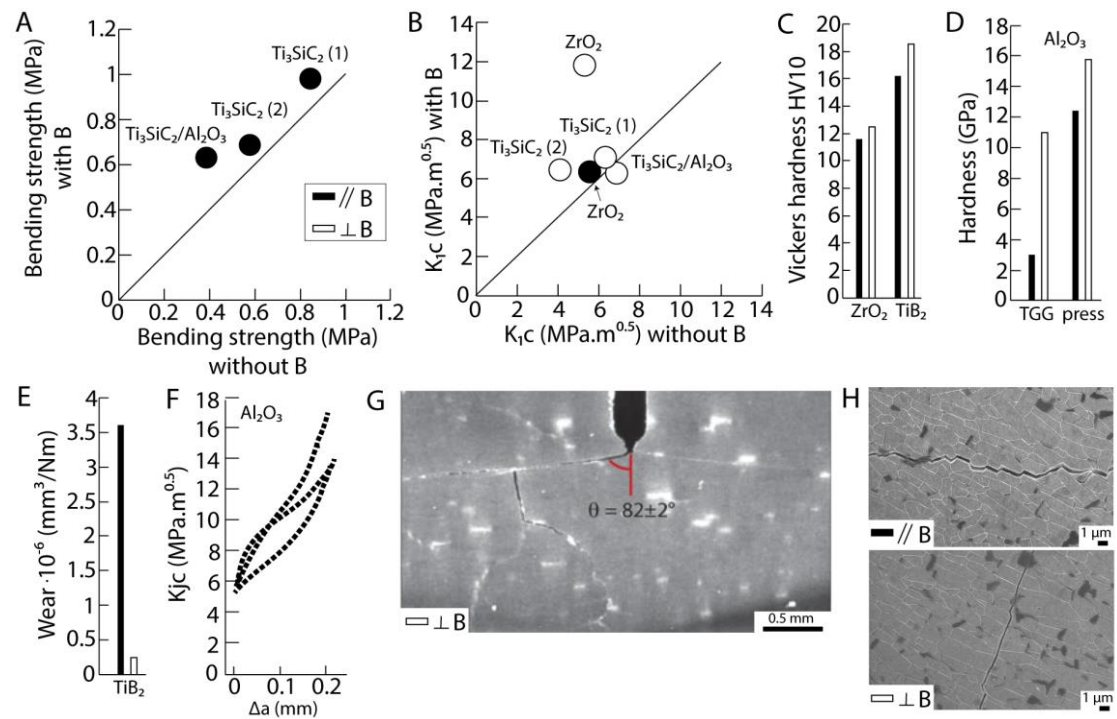

**Figure 7:** For all: lack indicates measurements parallel to the direction of the magnetic field imposed during fabrication and white, perpendicular. **(A)** Bending strength in slip casted samples prepared with magnetic field as a function of the bending strength of samples prepared without magnetic field. Data extracted from refs [72,109,110]. **(B)** Toughness at the crack initiation in samples prepared with magnetic field as a function

of the toughness in samples prepared without. Data extracted from refs [72,79,109,110]. **(C)** Vickers hardness. Data extracted from refs [74,79]. **(D)** Nanohardness in alumina samples prepared by magnetic slip casting and templated grain growth (TGG) and magnetic slip casting followed by hot pressing. Data extracted from refs [111,113]. **(E)** Wear. Data extracted from ref [74]. **(F)** Toughness $K_{JC}$ as a function of the crack length $\Delta a$ in a horizontally aligned sample. Data extracted from ref [61]. **(G)** Crack propagating in an aligned alumina ceramic in the perpendicularly to the texture. Reproduced with permission from [123]. Copyright 2020, Materials Research Society. **(H)** Electron micrographs showing cracks propagating parallel to the texture (top) and perpendicularly (bottom) in a SiC ceramic. Reproduced with permission from [82]. Copyright 2014, Elsevier Ltd and Techna Group S.r.l.

**5.2| Functional properties**

In addition to structural properties, functional properties are also enhanced by texturation. Electrical, thermal, piezoelectric and optical properties are summarized in the following (**Figure 8**).

Ceramics are generally dielectric materials with electrically insulating properties. Orienting the texture can create some anisotropy and increase the resistivity along the alignment direction in a variety of compositions (**Figure 8A**). However, as a dielectric, ceramics can get polarized under an electric field. When polarized, the electromagnetic energy can be dissipated by heat or vibrations causing dielectric loss. The inverse of this dielectric loss is quantified by the quality factor *Qf*. It was found that magnetic slip casting could create high anisotropy in the quality factor with a *Qf* value nearly 5 times higher parallel to direction of alignment than perpendicularly (**Figure 8B**). Similarly, piezoelectric properties could be enhanced by magnetic orientation, although this depends on the chemical composition (**Figure 8C**). The anisotropy was also found to depend on the quality of the texturation, with the composition PZT-A showing a Lotgering factor of 0.25 only after magnetic alignment and homogeneous piezoelectricity, whereas PZT-B had a value of 0.77 andanisotropic piezoelectric parameters [91].

Thermal properties of magnetically textured ceramics also exhibit anisotropy (**Figure 8D-F**). A higher thermal anisotropy was reported for $Si_3N_4$ and milder one for $Bi_2Te_3$-based ceramics (**Figure 8D**) [68,102,107]. The enhancement of anisotropy may depend on the composition but also on the microstructure of the sintered ceramics. Whereas the $Si_3N_4$ samples were constituted of elongated whiskers of 10 to 20 μm length and 1 to 2 μm diameter [102], the $Bi_2Te_3$ ceramics prepared had flake-shape grains of less than 5 µm diameter and 200 nm thickness [68] and the $Bi_{0.5}Sb_{1.5}Te_3$ had large grains of 30 µm diameter dispersed in a matrix of round shaped particles of ranging from 0.5 to 5 µm [107]. Indeed, larger grains size reduces the number of grain boundaries and interfaces, thus have a high thermal resistance. Thermoelectric properties have been measured on $Bi_2Te_3$ ceramics showing an enhancement of the Seebeck coefficient in magnetically textured samples (**Figure 8E**) [107]. This indicates that the orientation can increase the thermoelectric voltage generated by a temperature gradient across the sample. Also, thermal expansion can exhibit anisotropy. For example, in an $Al_2O_3$ ceramic with oriented plate-like grains, the relative width of the sample between 100 and 1100°C was reported to be nearly three times higher perpendicularly to the orientation as compared to the parallel direction (**Figure 8F**) [111]. This anisotropy is likely related to the anisotropy in elastic modulus that is higher along the direction of the alignment and thus resists the deformation. However, this anisotropy can be mitigated with an addition of an interlayer of BN between the $Al_2O_3$ grains.

Finally, magnetically textured ceramics have great potential for transparent ceramics (**Figure 8G-I**). For example, the transparency of $Al_2O_3$ was found to increase from 20 to 80% as the magnetic field and the alignment are increased (**Figure 8G**) [86]. As a result, the in-line transmittance at 640 and 420 nm has been reported to increase with the application of a magnetic field, in $Al_2O_3$ and LCO:Ce ceramics (**Figure 8H-I**) [86,88,116,125].

Thanks to the enhancement of structural and functional properties in ceramics prepared by magnetic slip casting, those new materials may find a broader range of applications. Next are discussed the potential and limits of magnetic slip casting.

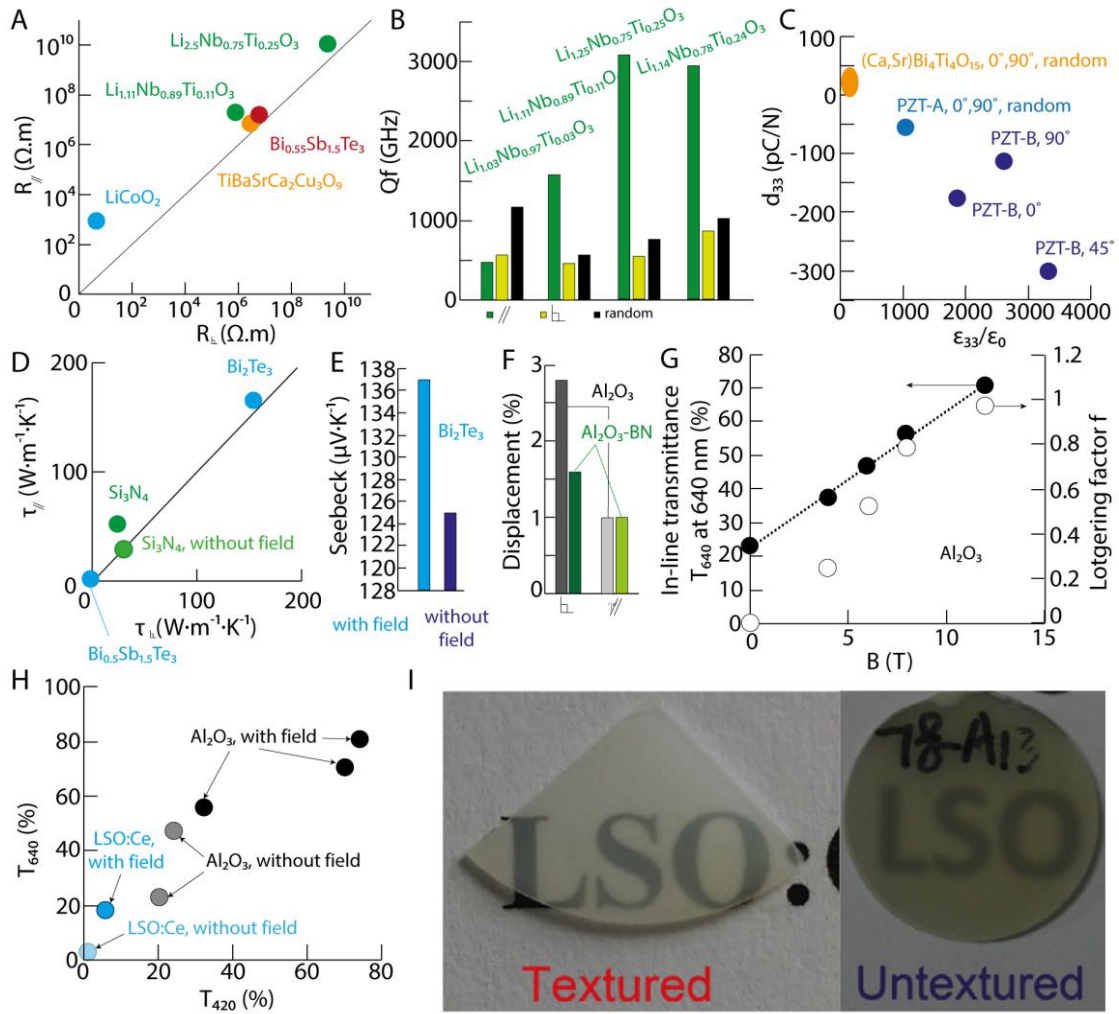

**Figure 8: (A)** Electrical resistance parallel to the magnetic orientation as a function of the resistance in the perpendicular direction. Data extracted from refs [67,68,78,107]. **(B)** Quality factor *Qf* as a function of the texture orientation for several ceramic compositions. Data extracted from ref [78]. **(C)** Piezoelectric coefficient $d_{33}$ and normalized permittivity $\frac{\varepsilon_{33}}{\varepsilon_0}$ for several ceramic compositions. PZT-A refers to Pb($Zr_{0.52}Ti_{0.48}$)$O_3$ and PZT-B to Pb($Zr_{0.23}Ti_{0.5}$($Ni_{0.33}Nb_{0.66}$)$_{0.27}$)$O_3$. Data extracted from refs [91,105,126]. **(D)** Thermal conductivity parallel to the orientation as a function of the perpendicular conductivity. Data extracted from ref [68,102,107]. **(E)** Seebeck coefficient of $Bi_2Te_3$ with and without magnetic alignment. Data extracted from ref [107]. **(F)** Relative expansion measured at 1100°C of $Al_2O_3$ and Al2O3-BN aligned ceramics, perpendicular and parallel to the alignment. Data extracted from ref [111]. **(G)** In line transmission of light at 640 nm as a function of the applied magnetic field for $Al_2O_3$. Data extracted from ref [86]. **(H)** In line transmission at 640 nm as a function of the transmission at 420 nm for several ceramics with and without magnetic alignment. Data extracted from refs [86,88,116,125]. **(I)** Optical images showing the transparency of a magnetically aligned and randomly aligned LSO:Ce ceramic. Reproduced with permission from [125]. Copyright 2017, Elsevier B.V.

## 6| POTENTIAL APPLICATIONS AND AREAS OF IMPROVEMENT

Magnetic slip casting has great potential for fabricating anisotropic materials with enhanced mechanical and functional properties, that are not typically easily realized in ceramics, such as toughness and transparency. This is made possible by applying remote magnetic fields and using adequate slurry compositions and sintering process. The resulting alignment and the angle of alignment obtained are much greater than compared with other techniques like tape casting or layer by layer deposition, where grains are aligned only in the horizontal direction ($\theta = 90°$, **Figure 9A**). Furthermore, micrometer to centimeter scale ceramics created using magnetic slip casting have been reported, in parallelepipedic and other shapes, and for a large variety of chemical compositions (**Table 4**). This shows great potential of the method for concrete applications. Although the specific processing parameters in terms of magnetic field, sample rotation, slurry composition, sintering process, etc. have to be optimized case-by-case, the alignment was found to increase during the densification by sintering. The correlation between alignment of the texture and the density of the ceramic is an additional advantage of the method, since a highly aligned and high density ceramic is generally desired (**Figure 9B**). Furthermore, the recent trend to move to lower magnetic fields strength further facilitate the implementation of the set up for complex and local control of the orientation [113].

Nevertheless, several areas of improvement remain. First, although alignment increases along with the density, the overall density achieved after magnetic orientation remains lower than in conventional ceramics. This is likely due to the slurries that are less concentrated to allow for the rotation of the grains and their alignment with the field. Also, in the case of mixtures of anisotropic and spherical powders, there might be a looser packing of the nanopowders around the anisotropic grains leading to closed nanopores [127]. To achieve the highest combination of alignment and density, uniaxial pressure can to be used (**Figure 9C**) [82]. However, this limits the texture to horizontal orientation. This is thus not favorable if vertical or local crystalline texture and grain orientation is desired, like in bio-inspired ceramics [61,113,127]. Secondly, it is still not clear how good the alignment should be to impact properties such as toughness. For example, ceramics with aligned structures but a misalignment of 30° prepared by hot pressing green bodies of anisotropic plate-like particles were found to display delayed fracture and a rising R-curve with a stress intensity factor $K_{jc}$ rising from 6.6 to 17.6 MPa.m$^{0.5}$ [128], whereas ceramics with only 5° misalignment prepared by tape-casting remained brittle without R-curve behavior [129]. The difference in interfacial properties between the grains is likely to be the key factor for the mechanical properties. It is thus difficult to decouple the influence of interfaces and grain sizes with the influence of the texture. Another question is how much anisotropy is desirable. Indeed, anisotropy may cause internal stresses to develop and could result in warping, distortion or cracking [127]. Finally, in view of practical applications, it remains a question of how to design the texture of ceramic parts to enhance its properties along one direction without weakening the properties in another direction. Most of materials are submitted to cyclic 3D mechanical or thermal stresses, and still need to perform their structural and functional roles.

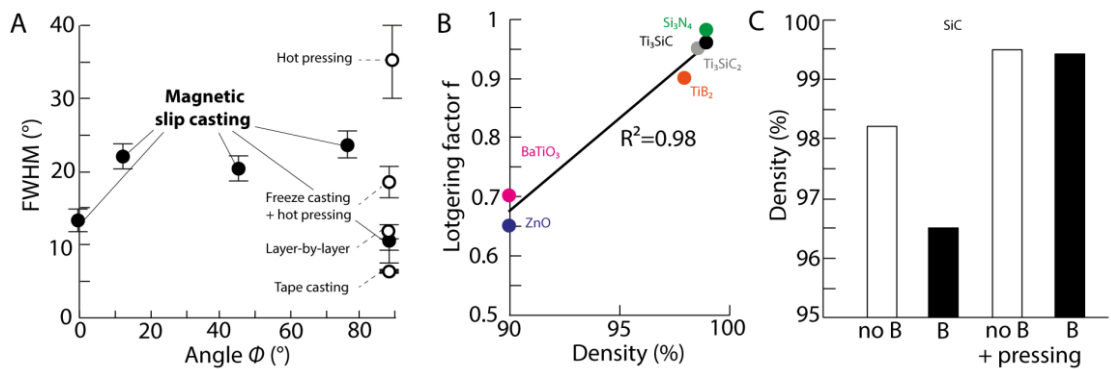

**Figure 9: (A)** Full width at half maximum (FWHM) of Al$_2$O$_3$ prepared by different methods as a function of the angle of alignment with respect to the vertical direction. Data extracted from refs [113,128,130–132]. **(B)** Lotgering factor as a function of the density of sintered ceramics prepared by magnetic slip casting. The line is a linear fit between the data points. Data extracted from refs [74,92,96,101,110,117]. **(C)** Density of SiC ceramics prepared with magnetic field (black) or without (white) and with conventional sintering or with pressure-assisted sintering. Data extracted from ref [82].

## 7| CONCLUSION

Magnetic slip casting is a process that creates anisotropy in ceramic materials by application of an external magnetic field during casting. This review presented the principles of slip casting, sintering, magnetic orientation, and the types of set-ups used. The resulting textures can be characterized by a variety of methods. Thanks to the controlled texture achieved, anisotropic properties could be achieved in a large pool of ceramic chemistries, with enhanced mechanical and functional properties along a preferential direction. Many questions still remain, but the results obtained so far demonstrate the potential of the method to create advanced ceramics that may exhibit toughness, transparency, or a combination, for example. Future research in the field is likely to tackle texture-properties relationships, control of local texture and density, combination with additive manufacturing and combination of magnetic alignment with magnetically-driven gradients. The ceramics to result would find applications in areas under high mechanical demands and stringent conditions of temperature, chemistry, etc. or areas where weight reduction while maintaining electrical and thermal properties are required.

## ACKNOWLEDGMENTS

Funding from the National Research Foundation, Singapore under the NRFF program (grant NRFF12-2020-0002) is acknowledged.